\newcounter{myctr}
\def\myitem{\refstepcounter{myctr}\bibfont\noindent\ifnum\themyctr>9\else\phantom{0}\fi\hangindent17pt\themyctr.\enskip}

\documentclass{ws-ijqi}
\usepackage{times}
\usepackage{graphicx}
\usepackage{color}
\usepackage{dcolumn}
\usepackage{bm}
\usepackage{subfigure}
\usepackage[T1]{fontenc}
\begin{document}

\markboth{Van Meter, Ladd, Fowler and Yamamoto} {Distributed Quantum
Computation Architecture using Semiconductor Nanophotonics}

\catchline{}{}{}{}{}

\title{DISTRIBUTED QUANTUM COMPUTATION ARCHITECTURE USING SEMICONDUCTOR NANOPHOTONICS}
\author{RODNEY VAN METER$^{1,*}$, %
        THADDEUS D. LADD$^{2,3}$, %
        AUSTIN G. FOWLER$^{4}$, \\
        and YOSHIHISA YAMAMOTO$^{2.3}$}
\address{%
$^1$Faculty of Environment and Information Studies,
    Keio University, \\
    5322 Endo, Fujisawa, Kanagawa, 252-8520, Japan\\
$^2$Edward L. Ginzton Laboratory, Stanford University,
    Stanford, CA, 94305-4088, USA\\
$^3$National Institute of Informatics, 2-1-2 Hitotsubashi,
  Chiyoda-ku, Tokyo-to 101-8430, Japan\\
$^4$Center for Quantum Computing Technology,
    University of Melbourne, Victoria 3010, Australia\\
 *Email: rdv@sfc.wide.ad.jp}
\maketitle


\begin{abstract}
  In a large-scale quantum computer, the cost of communications will
  dominate the performance and resource requirements, place many
  severe demands on the technology, and constrain the architecture.
  Unfortunately, fault-tolerant computers based entirely on photons
  with probabilistic gates, though equipped with ``built-in''
  communication, have very large resource overheads; likewise,
  computers with reliable probabilistic gates between photons or quantum
  memories may lack sufficient communication resources in the
  presence of realistic optical losses.  Here, we consider a
  compromise architecture, in which semiconductor spin qubits are coupled
  by bright laser pulses through nanophotonic waveguides and cavities
  using a combination of frequent probabilistic and sparse determinstic entanglement
  mechanisms.  The large photonic resource requirements
  incurred  by the use of probabilistic gates for quantum
  communication are mitigated in part by the potential
  high-speed operation of the semiconductor nanophotonic
  hardware.  The system employs topological cluster-state
  quantum error correction for achieving fault-tolerance.
  Our results suggest that such an
  architecture/technology combination has the potential to scale to a
  system capable of attacking classically intractable computational
  problems.
\end{abstract}


\section{Introduction}
\label{sec:Introduction}

Small quantum computers are not easy to build, but are certainly
possible.  For these, it is sufficient to consider the five basic
DiVincenco criteria\cite{divincenzo2000piq,divincenzo95:qc}: ability
to add qubits, high-fidelity initialization and measurement, low
decoherence, and a universal set of quantum gates.  However, these
criteria are insufficient for a large-scale quantum computer.
DiVincenzo's added two communications criteria --- the ability to
convert between stationary and mobile qubit representations, and to
faithfully transport the mobile ones from one location to another
and convert back to the stationary representation --- are also
critical, but so is gate speed (``clock rate''), the parallel
execution of gates, the necessity for feasible large-scale classical
control systems and feed-forward control, and the overriding issues
of manufacturing, including the reproducibility of structures that
affect key tuning
parameters~\cite{spiller:qip-intro-cp,van-meter:qarch-impli}. In
light of these considerations, the prospects for large-scale quantum
computing are less certain.

Advances in understanding what constitutes an attractive technology
for a quantum computer are married to advances in quantum error
correction.  These improvements include the theoretical thresholds
below which the application of quantum error correction actually
\emph{improves} the error rate of the
system~\cite{aharonov99:_threshold}, increases in the applicability of
known classical
techniques~\cite{ToddBrun10202006,bacon-2006,bacon05:_operator-self-qec,mackay2004sgc},
understanding of feasible implementation of error correcting
codes~\cite{steane02:ft-qec-overhead,steane02:_quant-entropy-arch,devitt04:_shor_qec_simul,copsey:q-com-cost,szkopek06,thaker06:_cqla,whitney:isca09},
design of error suppression techniques suited to particular
technologies or error
models~\cite{collin04:_nmr-like,vandersypen04:_nmr_techn,lidar98:dfs,lidar03:dfs-review,dur2007epa,evans-2007,fowler04:_qec_lnn,kitaev97:_quant},
advances in purification
techniques~\cite{bennett1996pne,cirac97:_distr_quant_comput_noisy_chann,dehaene2003lpp,dur2007epa,kruszynska2006epp,maneva2000itp,van-meter07:banded-repeater-ton},
and experimental advances toward
implementation~\cite{knill01:_bench,chiaverini04:_qec-realiz,pittman05:_optical-qec-demo}.
Among the most important, and radical, new ideas in quantum error
correction is ~\emph{topological quantum error correction} (tQEC), for
example \emph{surface
  codes}~\cite{dennis:topo-memory,raussendorf07:_topol_fault_toler_in_clust,kitaev2003ftq,freedman40tqc,raussendorf07:_2D_topo}.
These codes are attracting attention due to their high error
thresholds and their minimal demands on interconnect geometries, but
work has just begun on understanding the impact of tQEC on quantum
computer architecture, including determining the hardware resources
necessary and the performance to be expected~\cite{devitt:classproc2009,devitt08,divincenzo09,stock:170501}.

The effective fault tolerance threshold in tQEC depends critically
on the microarchitecture of a system, principally the set of qubits
which can be regarded as direct neighbors of each qubit. As
connectivity between qubits increases, both the operations required
to execute error correction and the opportunities for ``crosstalk''
as sensitive qubits are directly exchanged decline, allowing the
system to more closely approach theoretical limits.

Here, we argue that even for tQEC schemes that require only
nearest-neighbor quantum gates in a two-dimensional lattice geometry,
communication resources will continue to be critical.  We present an
architecture sketch in which efficient quantum communication is used
to compensate for architecture inhomogenities, such as physical qubits
which must be separated by large effective distances due to hardware
constraints, but also due to qubits missing from the lattice due to
manufacturing defects. Assuming a homogeneous architecture may be
acceptable for small-scale systems, but in order to create a system
that will grow to solve practical, real-world problems, distributed
computation and a focus on the necessary communications is
required. Further, our design explicitly recognizes that not all
communications channels are identical; they vary in the fidelity of
created entanglement and physical and temporal resources required.
This philosophy borrows heavily from established principles in
classical computer
architecture~\cite{hennessy-patterson:arch-quant4ed}. Classically,
satisfying the demands of data communication is one of the key
activities of system architects~\cite{dally04:_interconnects}. Our
design process incorporates this philosophy.

No computing system can be designed without first considering its
target \emph{workload} and \emph{performance
  goals}~\cite{hennessy-patterson:arch-quant4ed,van-meter:arch-dep-shor}.
The level of imperfection we allow for quantum operations depends
heavily on the application workload of the computer.  Our goal is
the detailed design (and ultimately implementation) of a large-scale
system: more than ten thousand logical qubits capable of running
$10^{11}$ Toffoli gates within a reasonable time (days or at most a
few months).  For example, such a system could factor a 2,000-bit
number using Shor's algorithm~\cite{shor:factor}.  This choice of
scale affects the amount of error in quantum operations that we can
tolerate.  Steane analyzes the strength of error resilience in a
system in terms of $KQ$, the product of the number of logical qubits
in an application ($Q$) and the depth (execution time, measured in
Toffoli gate times) of the application
($K$)~\cite{steane02:ft-qec-overhead}.  Our goal is to tune the
error management system of our computer to achieve a logical error
per Toffoli gate executed of $p_L \ll 1/KQ$, with $KQ \sim
10^{15}$~\cite{van-meter06:thesis}.

Under most realistic technological assumptions, the resources
required to reach adequate $KQ$ values are huge. Nearly all proposed
matter qubits are at least microns in size, when control hardware is
included.  For chip-based systems, a simple counting argument
demonstrates that more qubits are required than will fit in a single
die, or even a single wafer. This argument forces the implementation
to adopt a distributed architecture, and so we require that a useful
technology have the ability to entangle qubits between
chips~\cite{van-meter06:thesis,yepez01:_type_ii}.

As an example architecture supporting rich communications, we are
designing a device based on semiconductor nanophotonics, using the
spin of an unpaired electron in a semiconductor quantum dot as our
qubit, with two-qubit interactions mediated via cavity QED.  We plan
to use tQEC to manage run-time, soft faults, and to design the
architecture to be inherently tolerant of fabricated and grown
defects in most components.

Our overall architecture is a \emph{quantum multicomputer}, a
distributed-memory system with a large number of nodes that
communicate through a multi-level interconnect.  The distributed
nature will allow the system to scale, circumventing a number of
issues that would otherwise place severe constraints on the maximum
size and speed of the system, hence limiting problems for which the
system will be suitable.

Within this idiom, many designs will be possible.  The work we present
here represents a solid step toward a complete design, giving a
framework for moving from the overall multicomputer architecture
toward detailed node design.  We can now begin to estimate the actual
hardware resources required, as well as establish goals (such as the
necessary gate fidelity and memory lifetimes) for the development of
the underlying technology.

Section~\ref{sec:background} presents background on the techniques for
handling of errors in a quantum computer that we propose to use.
Section~\ref{sec:hardware} qualitatively presents our hardware
building blocks: semiconductor quantum dots, nanophotonic cavities and
waveguides, and the optical schemes for executing
gates. Section~\ref{sec:description} presents a qualitative
description of the resources employed in the complete system.  In
particular, it describes how some quantum dots, used for
communication, are arranged for deterministic quantum logic mediated
by coupled cavity modes, while other quantum dots are indirectly
coupled via straight, cavity-coupled waveguides for
purification-enhanced entanglement creation.  Long columns of these
basic building blocks span the surface of a chip, and many chips are
coupled together to create the complete multicomputer.  Preliminary
quantitative resource counts appear in section~\ref{sec:results}.

\section{Multi-level Error Management}
\label{sec:background}

A computer system is subject to both \emph{soft faults} and \emph{hard
  faults}; in the quantum computing literature, ``fault tolerance''
refers to soft faults.  A soft fault is an error in the operation of a
normally reliable component.  Soft faults can be further divided into
errors on the quantum state (managed through dynamically-executed
quantum error correction or purification), and the loss of qubit
carrier (e.g., loss of a photon, ion or the electron in a quantum dot,
depending on the qubit technology).  Qubit loss may be addressed by
using erasure codes, or, in the case of tQEC, through special
techniques for rebuilding the lattice state~\cite{stace:200501}.  In
this section, we introduce our approach to managing these multiple
levels of errors, which will be further developed in the following
sections.

\subsection{Defect Tolerance and Quantum Communication}

Hard faults are either manufactured or ``grown'' defects (devices that
stop working during the operational lifetime of the system).  With
adequate hardware connectivity, flexible software-based assignment of
roles to qubits will add hard fault tolerance, allowing the system to
deal with both manufactured and grown defects.

The percentage of devices that work properly is called the
\emph{yield}.  In our system, most of the components are expected to
have high yields, but the quantum dots themselves will likely have
low yields, at least in initial fabrication runs and possibly in
ultimate devices.  These faults occur in part due to the difficulty
of growing optically active quantum dots in prescribed locations,
but more due to the difficulty of assuring each dot is appropriately
charged and tuned near the optical wavelength of the surrounding
nanophotonic hardware, to be further discussed in
Sec.~\ref{sec:gates}.

The presence of hard faults means that the connectivity of the quantum
computer begins in a random configuration, which we can determine by
device testing. As a result, the architecture will have an
inhomogeneous combination of high-fidelity connections where pairs of
neighboring qubits are good and low-fidelity connections between more
distant qubits.  To compensate for the low-fidelity connections, we
choose to use \emph{entanglement purification} to bring long-distance
entangled-states up to the fidelity we desire for building our
complete tQEC lattice. This choice means that the system will
naturally use many of the techniques developed for quantum
repeaters~\cite{briegel98:_quant_repeater,dur2007epa,van-meter07:banded-repeater-ton},
and portions of the system will require similar computation and
communication resources, used in a continuous fashion.  Details of
these procedures are presented in Sec.~\ref{sec:description}.

\subsection{Topological Fault Tolerance}
\label{sec:topo}

On top of purified states, we employ \emph{topological error
  correction} (tQEC),
\cite{dennis:topo-memory,raussendorf07:_topol_fault_toler_in_clust,kitaev2003ftq,freedman40tqc},
in particular the two-dimensional scheme introduced by
Raussendorf and
Harrington\cite{raussendorf07:_2D_topo,fowler2008htu,fowler-2008}.
In this scheme, the action of the quantum computer is the
sequential generation and detection of a cluster state, and
error correction proceeds by checking against expected quantum
correlations for that state.  Logical qubits are defined by
deliberately altering these correlations at a pair of
boundaries in an effectively three-dimensional lattice of
physical qubits.  These boundaries may be the extremities of
the lattice or holes\footnote{These holes are
  commonly called ``defects'' in the topological computing literature,
  as they are similar to defects in a crystal; in this paper, we
  reserve the term ``defect'' for a qubit that does not function
  properly, i.e. a manufacturing defect.} of various shapes ``cut''
into the lattice by choosing not to entangle some qubits.  The
qubits in the interior of the lattice have their state tightly
constrained, whereas pairs of boundaries are associated with a
degree of freedom that is used as the logical qubit.

The simplicity of the gate sequences used to constrain the qubits in
the lattice interior and the independence of these gate sequences on
the size of the system are directly responsible for tQEC's high
threshold error rate of approximately 0.8\% for preparation, gate,
storage and measurement
errors~\cite{raussendorf07:_topol_fault_toler_in_clust,wang2009ter},
the highest threshold found to date for a system with only nearest
neighbor interactions.

\begin{figure}
\centerline{\hbox{
\includegraphics[width=10cm]{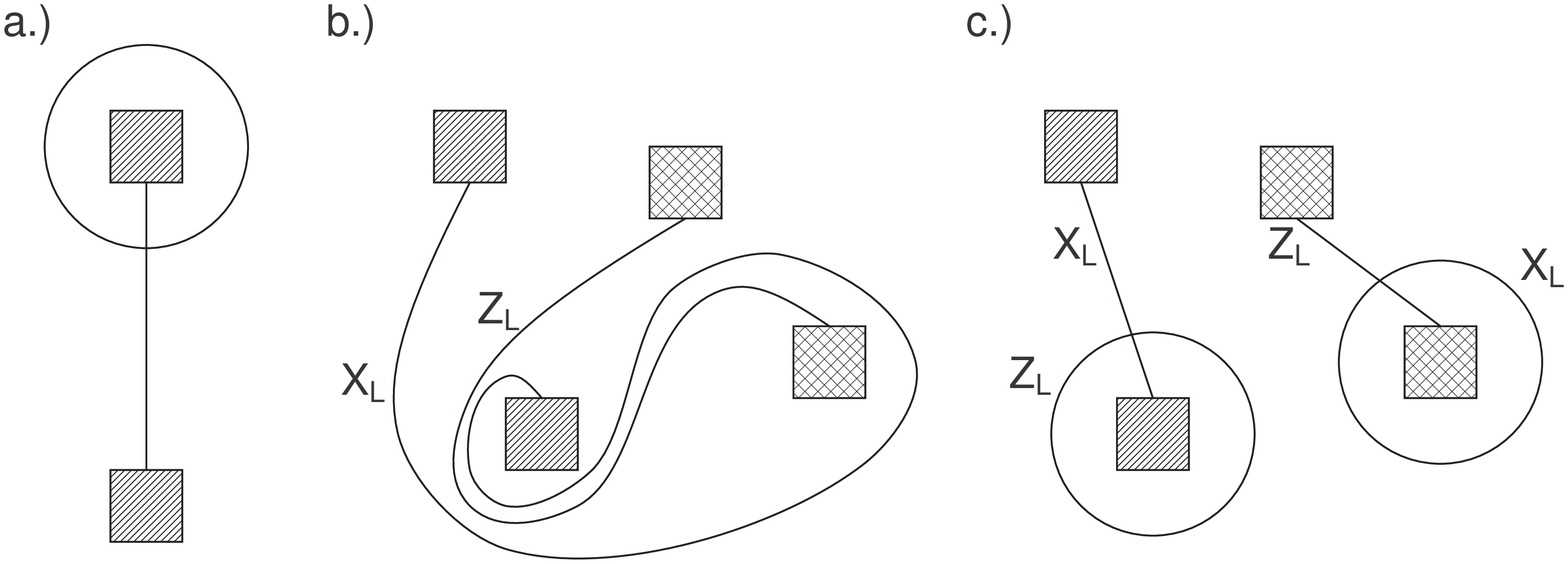}}}
\caption{Logical qubits in topologically error-corrected systems are
  represented by unentangled ``holes'' in a high-entangled cluster
  state on a lattice.  The lattice itself is not shown; the squares
  represent the holes.  a.) A single logical qubit is associated with
  two holes.  Logical operators are rings and chains of single qubit
  operators.  b.) Moving holes around one another by changing the
  error correction circuits on the boundary of holes results in the
  deformation and ultimately braiding of logical operators.  c.)
  Equivalent form of the braided logical operators after pinching
  together sections, and thus cancelling these sections, to form
  disjoint rings and chains.  The mapping of logical operators
  represents logical CNOT with the left logical qubit as control.}
\label{fig:braiding}
\end{figure}

In 2-D, we choose to make holes that are squares of side length $d$.
Logical operators take the form of rings and chains of single-qubit
operators --- chains connect pairs of holes, rings encircle one of
the holes.  If we associate $X_L$ with chains and $Z_L$ with rings
(or vice versa), it can be seen that these operators will always
intersect an odd number of times ensuring anticommutation.  Braiding
holes around one another can implement logical CNOT, as shown in
Figure~\ref{fig:braiding}.

tQEC offers important architectural advantages over other
error-suppression schemes, such as concatenated codes.  Most
importantly, unlike tQEC,
many concatenated codes lose much of their effectiveness when
long-distance gates are precluded by the underlying technology. In
addition, the amount of error correction applied in tQEC can be
controlled more finely than with concatenated codes, which have a
property that every time an additional level of error correction is
used, the number of physical qubits grows by at least an order of
magnitude. tQEC's error-protection strength, in contrast, improves
incrementally with each additional row and column added to the
lattice.

Logical errors are exponentially suppressed by increasing the
circumference and separation of holes.  This can be inferred
directly from Figure~\ref{fig:braiding} --- the number of physical
qubit errors required to form an unwanted logical operation grows
linearly with circumference and separation.  The threshold error
rate $p_{th}$ is defined to be the error rate at which increasing
the resources devoted to error correction neither increases nor
decreases the logical error
--- the error rate at which the errors corrected are balanced by the
errors introduced by the error correction circuitry.  Assuming a
hole circumference and separation of $4d$, for physical error rates
$p<p_{th}$, error suppression of order $O((p/p_{th})^{\alpha d})$
will be observed.  The factor $\alpha$ depends on the details of the
error correction circuits.  Assuming the error correction circuits
do not copy single errors to multiple locations, $\alpha\sim 2$ as a
circumference of $4d$ implies that a chain of approximately $2d$
errors can occur before our error correction system will mis-correct
the state and give a logical error.

Related tQEC schemes exist in 3-D and 2-D
\cite{raussendorf07:_topol_fault_toler_in_clust,fowler2008htu,fowler-2008,raussendorf07:_2D_topo}.
The 3-D scheme makes use of a 3-D cluster state and the
measurement-based approach to computing --- all qubits are
measured in various bases, and measurement results processed to
determine both the bases of future measurements and the final
result of the computation. This approach is well-suited to a
technology with short-lived qubits (e.g., photons, which are
easily lost) or slow measurement.  The 2-D scheme requires a
2-D square lattice of qubits that are not easily lost plus fast
measurement.  Given these two properties, the threshold is
slightly higher than the 3-D case and certain operations, such
as logical measurement, can be performed more quickly.  Barring
these minor caveats, the 2-D scheme is a simulation of the 3-D
scheme, in which one dimension of the 3-D lattice becomes time.

\subsection{Logical Gates in Topological Error-Corrected Systems}

When making use of topological error correction, only a small
number of single logical qubit gates are possible --- namely
$X_L$, $Z_L$ and logical initialization and measurement in
these bases. Logical initialization and measurement in the
$X_L$ and $Z_L$ bases can be implemented using initialization
and measurement of regions of single qubits encompassing the
defects in the $X$ and $Z$ bases. The only possible multiple
logical qubit gate, logical CNOT, can be implemented by
braiding the correct type of defects in a prescribed manner as
shown in Figure~\ref{fig:braiding}.  This set of gates is not
universal.

To achieve universality, rotations by $\pi/2$ and $\pi/4$ around the
$X_L$ and $Z_L$ axes can be added to the logical gate set.  These
gates, however, require the use of specially-prepared $S$ states
where $|S\rangle = |0\rangle+e^{i\theta}|1\rangle$,
$\theta=\pi/2,\pi/4$. Fault-tolerant creation of the $S$ states
involves use of the concatenated decoding circuits for the 7-qubit
Steane code and 15-qubit Reed-Muller code respectively to distill a
set of low-fidelity $S$ states into a single higher-fidelity one.
Convergence is rapid --- if the input states have average
probability of error $p$, the output states will have error
probabilities of $7p^3$ and $35p^3$
respectively~\cite{raussendorf07:_topol_fault_toler_in_clust}.

This implies that for most input error rates, two levels of
concatenation will be more than sufficient.  Nevertheless, this
still represents a large number of logical qubits, implying the
need for $S$ factories throughout the computer and the
dedication of most of the qubits in the computer to generate
the necessary $S$ states at a sufficient rate.  This will
impact the resource counting for our target application, as we
discuss in Section~\ref{sec:results}.

When using an $S$ state, the actual gate applied will be a random
rotation by either $+\theta$ or $-\theta$.  Error corrected logical
measurement must be used to determine which gate was applied and
hence whether a corrective $2\theta$ gate also needs to be applied.
If $2\theta=\pi/2$, the correction must be applied before further
gates are applied, introducing a temporal gate ordering.  This time
ordering prevents arbitrary quantum circuits involving non-Clifford
group gates being implemented in constant time.

\section{Hardware Elements}
\label{sec:hardware}

In considering the harware in which to implement this
architecture, by far the most important pending question is the
choice of quantum dot type, which will also determine the
semiconductor substrate and operational wavelengths.

\subsection{Quantum Dots}

The best type of quantum dot to employ remains an open
question. Charged, self-assembled InGaAs quantum dots in GaAs
are appealing due to their high oscillator strength and near-IR
wavelength.  These dots have been engineered into cavities in
the strong coupling regime~\cite{reithmaier2004strong} and recent
experiments have demonstrated complete ultrafast optical
control of a single electron spin qubit trapped in the
dot\cite{awschsingle,pressnature}.  However, it is challenging
to make high-yield CQED devices from these dots due to their
high inhomogeneous broadening and the challenges of site
selectivity, although progress continues in designing tunable
quantum dots\cite{forchel_tuning,vuckovic_phaseshift} in
prescribed locations\cite{forchel_alignment}. Sufficient
homogeneity for a scalable system, however, may require a more
homogeneous kind of quantum dot, such as those defined by a
single donor impurity and its associated donor-bound-exciton
state. Donor-bound excitons in high quality silicon and GaAs
are remarkably homogeneous, both in their optical transitions
and in the Larmor frequencies of the bound spin providing the
qubit.  However, the isolation of single donors in these
systems has been challenging.  Donor impurities in silicon
would seem almost ideal, since isotopic purification can give
long spin coherence times\cite{tlar03}  and extremely
homogeneous optical transitions\cite{thewalt}, but optical
control in this system is hindered by silicon's indirect
band-gap.  A II-VI semiconductor such as ZnSe may provide a
nearly ideal compromise -- single fluorine impurities in ZnSe
have been isolated, shown to have a comparable oscillator
strength to quantum dots, and incorporated into
microcavities\cite{ZnSelasing}. Recently, sufficient
homogeneity has been available to observe interference from
photons from independent devices\cite{znse_preprint}.  However,
this system comes with its own challenges, such as the less
convenient blue emission wavelength. Nitrogen-Vacancy centers
in diamond\cite{dutt_science,diamond_CPT,diamond_ultralongT2}
have also attracted heavy attention recently, but the diamond
substrate remains a challenging one for implementing the
nanophotonic hardware that supports the quantum computer.

Regardless of the type of quantum dot, there are several common
physical features which are to be employed for quantum information
processing.  The dot has a two-level ground state, provided by the
spin of trapped electrons in a global applied magnetic field.  This
spin provides the physical qubit.  The dot also has several optical
excited states formed from the addition of an exciton to the dot.  One
of these excited states forms an optical $\Lambda$-system with the two
ground states, allowing not only single qubit control via stimulated
Raman transitions\cite{clark2006qcb}, but also selective optical phase
shifts of dispersive light\cite{hybridnjp} (to be discussed in
Sec.~\ref{sec:gates}) or state-selective
scattering\cite{early_cirac_entanglement,childress05:_ft-quant-repeater,wv06}. These
enable several possible means to achieve entanglement mediated by
photons.

\subsection{Nanophotonics}

The quantum dots will be incorporated in small cavities to
enhance their interaction with weak optical fields.  Cavities
may be made from a variety of technologies, including photonic
crystal defects and microdisks.  Here, we will focus on
suspended microdisk cavities.

The small microdisks are in turn coupled to larger waveguides
arranged as disks, rings, or straight ridges, which carry
qubit-to-qubit communication signals.  These waveguides can be
ridges topographically raised above the chip surface, or
line-defects in photonic crystals. Our present focus is on
ridge-type waveguides. Waveguides are well-advanced and
relatively low-loss, although it is best to make the waveguides
as straight as possible, and to avoid crossing two waveguides
in the floor plan. Silicon at telecom wavelengths, for example,
makes a good waveguide for our purposes, as it is almost
transparent to $1.5~\mu$m light, with a loss of about 0.1dB/cm.
The coherent processing of single photons in on-chip waveguides
has recently been well demonstrated for ridge-type silica
waveguides\cite{LOQC_chip}.

The ``no crossing waveguides'' restriction is one of the two
key issues driving device layout.  The other is the need to
route signals to more than one possible destination, for which
high-speed, low-loss optical switching is required. Good
optical switches are difficult to build: many designs have poor
transmission of the desired signals and poor extinction of the
undesired ones, and tend to be large and slow.  In our
architecture, we focus on microdisk-type or microring-type
add/drop filters.  In suspended silica systems, these switches
have been shown to have insertion losses as low as 0.001 dB for
the ``bus'' when the microdisk is off-resonant; optical loss
from the bus to the drop port can be as low as 0.3 dB when the
system is resonant\cite{PhysRevLett.92.253905}.  On-chip
switches in semiconductor platforms do not typically feature
such nearly ideal behavior but continue to improve.  For
example, $40~\mu$m by $12~\mu$m multi-ring add-drop switches
with a loss of a few dB were recently demonstrated in a silicon
platform~\cite{vlasov08:_nano-optical-switch}.

We need to individually control the resonance of every optical
microdisk in the circuit; these microdisks provide the add/drop
switches and qubit-hosting cavities.   Ultimately, it is the ability
to rapidly move these microdisk resonators into and out of
near-resonance with the waveguided control light that provides the
quantum networking capability.  A candidate method for this is to
employ the optical nonlinearity of the semiconductor substrate. A
strong, below-gap laser beam focused from above onto one of the
cavities will shift its index of refraction through a combination of
heating, carrier creation, and intrinsic optical
nonlinearities~\cite{clark2006qcb}.  The laser pulses for this may
be carried through free space from a micromirror
array~\cite{kim05:_system}.

To complete the architecture, we will also need mode-locked lasers for
single-qubit control, modulated CW-lasers for quantum non-demolition
(QND) measurements as well as deterministic and heralded entanglement
gates, and photodiodes to measure the intensity of the control
light. Lasers and photodiodes are expensive in both space and
manufacturing cost, so an ideal system will be carefully engineered to
minimize the number required.  Mode-locked lasers with repetition
frequency tuned to the Larmor frequency of spin qubits will be used
for fast single-qubit rotations~\cite{clark2006qcb}.  These lasers may
be directed by the same micromirror used for switching. More slowly
modulated single-frequency lasers will be used for qubit
initialization, measurement, and entanglement operations. These lasers
may be incorporated into the chip, or injected via a variety of
coupling technologies.  The photodiodes are intended to measure
intensity of pulses with thousands to millions of photons, rather than
single-photon counting, which allows the possibility of fast, on-chip,
cavity-enhanced photodiodes; however, off-chip detectors may be more
practical depending on the semiconductor employed.

These resources are crucial, as they are needed for every
single-qubit measurement and heralded entangling operation. These
operations dominate the operation of a cluster-state-based quantum
computer.  However, these same technologies are evolving rapidly for
classical optoelectronic interconnects, and are expected to continue
to improve in coming years.

\subsection{Executing Physical Gates}
\label{sec:gates}

Four types of physical gates are employed in this architecture.

The first type of gate is arbitrary single qubit rotations, which may
be performed efficiently using picosecond pulses from a semiconductor
mode-locked laser with pulse repetition frequency tuned to the qubit's
Larmor frequency~\cite{clark2006qcb,pressnature}.  A cavity is not
needed for this operation, and the pulses used are sufficiently far
detuned from the qubit and the cavity resonance that the cavity plays
little role.  The phase and angle of each rotation is determined via
switching pulses through fixed delay routes, as described in
Ref.~\refcite{clark2006qcb}.  The performance of this gate is limited
by spurious excitations created in the vicinity of the quantum dot by
the pulse~\cite{clarkpreprint} and not by optical loss or other
architectural considerations.

The next type of gate is the quantum-non-demolition QND
measurement of a single qubit.  This gate is critical, since
the initialization and measurement of every qubit is very
frequent in our tQEC architecture, and the QND gate allows
both.  A QND measurement makes use of the optical microcavity
containing the dot, and operates with the cavity well detuned
from the dot's optical transitions.  In such a configuration,
an optical transition to one qubit ground state may present a
different effective index of refraction for a cavity mode than
the optical transition to the other qubit ground state.  This
results in a qubit-dependent optical phase shift of a slow
optical pulse coupled in and out of the waveguide.  This
optical pulse may then be mixed with an unshifted pulse from
the same laser to accomplish a homodyne measurement of the
phase shift.  In one variation of this scheme, this phase is
detected as a change in the polarization direction of a
linearly polarized optical probe beam; this has been
demonstrated for quantum dots both with\cite{awschalom_dps} and
without\cite{imamoglu_dps} a microcavity; larger phase shifts
have also been observed in neutral dots in improved photonic
crystal cavities\cite{ilya}.  Simulations indicate that pulses
with a timescale of about 100~ps may be used for this
gate~\cite{hybridnjp,clark2006qcb}.

These first two gate types are single-qubit gates.  For generating
entanglement between distant qubits, two further gates are employed:
a deterministic, nearest-neighbor gate, and a non-deterministic gate
for heralded entanglement generation for distant qubits.

The deterministic, nearest-neighbor gate will be mediated by a common
microdisk mode connecting the cavities joining nearby qubits. The
phase or amplitude of this cavity mode may be altered by the state of
the qubits with which it interacts, which in turn changes the phase or
population of those qubits.  The gate is achieved by driving the
coupled cavity mode with one or more appropriately modulated optical
pulses from a CW laser. The light is allowed to leak out of the cavity
and may then be discarded.  The amplitude version of such a gate was
proposed in 1999 by Imamoglu et al.\cite{IABDLSS99}, and may be viewed
as a pair of stimulated Raman transitions for two qubits driven by two
CW lasers and their common cavity mode. This gate is known to require
high-$Q$ cavities.  The phase version of this gate, described in
Ref.~\refcite{PhysRevA.70.052320}, is an adaptation of the ``qubus''
gates proposed by Spiller et al. in 2006\cite{qubus}; more detailed
design and simulation of this gate in the present context is in
progress\cite{lyip}.

If such deterministic gates are available, one may naturally
ask whether a fully two-dimensional architecture of coupled
qubits is more viable than the communication-based architecture
we present here.  Indeed, if truly reliable cavity QED systems
can be developed in the large-scale, deterministic
photonic-based gates\cite{dk04} may enable highly promising
single-photon-based architectures for tQEC\cite{modulePRA}.
However, the devices that will enable deterministic CQED gates
in solid-state systems are unlikely to be fully reliable.

In particular, high-fidelity deterministic gates require extremely
low optical loss between qubits, and therefore cannot easily survive
coupling to straight waveguides or to other elements in the photonic
circuit such as switches and fibers. For generating entanglement
through these elements, stochastic but heralded entanglement schemes
are used, similar to gates in linear optics except with physical
quantum memory.  Combined with local single-qubit rotations, QND
measurements, and deterministic nearest-neighbor gates, this
heralded entanglement allows quantum teleportation. Heralded
entanglement is the bottleneck resource in quantum wiring. Heralded
entanglement gates come in several flavors, but fortunately each
type requires the same basic qubit and cavity resource; they vary in
the strength of the optical field used and the method of optical
detection. Which type to employ depends on the amount of loss
between the qubits to be entangled.

For qubits with relatively low loss between them, such as those
coupled to a common waveguide without traversing to the drop
port of a switch, so-called ``hybrid'' schemes are
attractive\cite{hybridprl,hybridnjp}. In these schemes, the QND
measurement discussed above is extended to two qubits,
distinguishing odd-parity qubit subspaces from even-parity
states.
For some detection schemes, such as $x$-homodyne detection, this
parity gate may be deterministic, up to single-qubit operations which
depend on measurement results~\cite{PhysRevA.71.060302,munro05:_weak}.
If such parity gates are available, ``repeat-until-success'' schemes
for quantum computation are very
attractive~\cite{PhysRevA.71.060310}, and have been proposed for use
in multicomputer-like distributed
systems~\cite{lim05:_repeat_until_success}.  However, if weak CQED
nonlinearities are employed with lossy waveguides, these detection
schemes fail~\cite{hybridprl,hybridnjp}.  In this case, $p$-homodyne
detection may still show strong performance, but the parity gate is
incomplete.  The heralded measurement of an odd-parity state may
project qubits into an entangled state with probability $\simeq50$\%, but
when this fails no entanglement is present.  As in schemes using
linear optics, this allows probabilistic quantum logic.  With the
addition of an extra ancilla qubit, this partial parity-gate may be
combined into a probabilistic CNOT gate for entanglement purification.

This scheme is attractive due to its use of relatively bright laser
light and near ideal probability of successful heralding.  However, it
is strongly subject to loss, as has been discussed
previously~\cite{hybridnjp}.  More complex measurement schemes may
improve the fidelity of such gates at the expense of their probability
of heralding a success~\cite{peter_USD}. For very lossy connections,
the number of photons in the optical pulse might be reduced to an
average of less than one photon, in which case single-photon
scattering
schemes~\cite{early_cirac_entanglement,childress05:_ft-quant-repeater,wv06}
would be employed. These schemes succeed much more infrequently, as
they rely on the click of a single photon detector projecting the
combined qubit/photon system into one where no photons were lost, a
possibility whose probability decreases with loss.  Here, we consider
only many-photon qubus gates using homodyne detection as discussed in
Ref.~\refcite{hybridnjp}; we compensate for different connections with
different loss rates only by changing the intensity of the optical
pulses employed, whose optimum varies with loss.  The detection scheme
remains constant across the architecture.

Although proposals for nonlocal, deterministic gates exist, their
performance is always hindered by optical loss.  This is an
inevitability: if photons are mediating information between qubits,
the loss of those photons into the environment inevitably reveals
some information about the quantum states of the qubits, causing
decoherence.  A well-designed photon-mediated architecture should
use a hierarchy of photon-mediation schemes to provide
high-success-probability gates at low distances and highly
loss-tolerant gates at higher distances, and the qubus mechanisms
allow some degree of hierarchical tuning without adding extra
physical resources.

In the present discussion, we discuss performance entirely in
terms of optical loss.  Photons may be lost in waveguides, from
cavities, from the cavity-waveguide interfaces, and from
spontaneous emission.  An approximation of the amount of
decoherence-causing loss at a quantum-dot-loaded cavity and
cavity/waveguide interface, when running hybrid CQED-based
gates optimally, is the inverse of the cooperativity factor
$C$~\cite{hybridnjp}. This factor arises from the ratio of
spontaneous emission into a cavity mode (assumed to be
overcoupled to the waveguide) to spontaneous emission into
other modes.  It scales as the quality factor of the cavity
divided by its mode volume, so the cavities containing qubits
are designed small to maximize this factor. When we discuss
qubit-to-qubit optical loss, this loss should be considered as
the linear loss in the waveguide connecting the qubits plus
about $C^{-1}$. Cooperativity factors between self-assembled
quantum dots and the whispering gallery modes of suspended
microdisks have been shown to approach
100~\cite{Gerard05a,koseki09:mono-integ}, corresponding to a
cavity-induced loss limit of 0.04 dB.

\section{Architecture: Layout and Operational Basics}
\label{sec:architecture}
\label{sec:description}

In this section, we qualitatively describe our architecture and its
operation.  Many of the design decisions described here will be
justified numerically in Section~\ref{sec:results}.

\subsection{Architecture Axes}

\begin{figure}
\centering
\includegraphics[width=0.8\columnwidth]{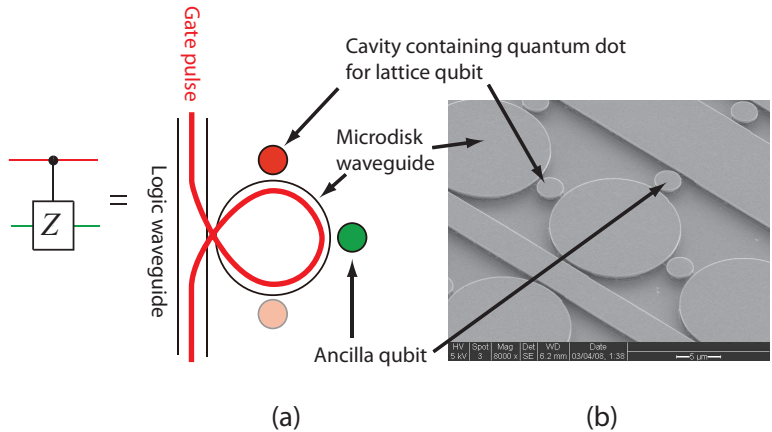}
\caption{(a) Layout and pulse path for executing a local,
  high-fidelity controlled-Z gate.  An optical pulse couples from the
  straight waveguide to the microdisk waveguide; the two qubits of
  interest are introduced to the logic gate by bringing their cavities
  into resonance with the optical pulse.  (b) Scanning electron
  micrograph of a non-functional demonstration device, fabricated in
  GaAs with (unshown) InAs quantum dot layer.  The structures are
  underetched following the methods presented in
  Ref.~93.}
\label{fig:local-cz}
\end{figure}

The basic structural element of our system is one-dimensional: a
waveguide with a tangent series of microdisks, each connected to one
or more smaller microdisks containing quantum dots, as in
Fig.~\ref{fig:local-cz}.  The shared bus nature of a single waveguide
offers the advantage that the qubit at one end can communicate quickly
and easily with the qubit at the other end; this long-distance
interaction has the potential to accelerate some algorithms and aids
in defect tolerance, as we will show below. However, that shared
nature makes the bus itself a performance \emph{bottleneck} in the
system, as contention for access to the bus and the measurement device
forces some actions to be
postponed~\cite{van-meter07:_distr_arith_jetc}.

This limitation on concurrent operation makes it natural to consider
using multiple columns.  Columns are connected by teleportation, aided
by heralded entanglement and purification.  The resulting structure,
developed in Figures ~\ref{fig:local-cz} to \ref{fig:qmc}, is a set of
many columns, defined by long, vertical waveguides, interspersed with
smaller, circular and oval waveguides, and qubits in cavities
tangential to the waveguides. The vertical waveguides are of two
types: \emph{logic} waveguides, which are used to execute operations
between qubits within one column, and \emph{teleportation} waveguides,
which are used to create and purify connections between columns within
a single chip or between chips.  The small, colored circles represent
the smallest microcavities containing quantum-dot qubits. The
different colors represent different roles for particular qubits,
which we describe in Section~\ref{sec:qubit-roles}.  The teleportation
columns do not use the smaller, higher-$Q$ circular waveguides to
couple qubits deterministically.  Instead, as in
Figures~\ref{fig:cnot-purification} and \ref{fig:cnot-cluster}, they
use larger racetrack-shaped waveguides that can support a larger
number of qubits which are only stochastically entangled, called
transceiver qubits.  The qubits along one racetrack can be used to
purify ancilla qubits, allowing us to connect qubits in potentially
distant parts of the chip, or to connect to off-chip resources.

The architecture in Fig.~\ref{fig:qmc} is designed to minimize both
the length of waveguides and the number of switches traversed by
pulses carrying quantum information.  Note that signals introduced
onto the waveguide snaking through the chip will not be perfectly
switched into the detectors, implying some accumulated noise; however,
this effect can be mitigated with appropriate detector time binning
and sufficiently large microdisk $Q$-factors in the switches.

A single node has two axes of growth.  The length of a logical
waveguide column and the number of columns provide the basic
rectangular layout, which will have some flexibility but is
ultimately limited by the size of chip that can be practically
fabricated, packaged and used.  To give a concrete example, if
we set the vertical spacing of the red lattice qubits to
$50~\mu$m and the column-to-column spacing to $100~\mu$m, 100
qubits in each vertical column and 100 columns will result in
the active area of the chip being 5~mm by 10~mm.

A third axis of growth is the number of chips that are
connected into the overall system -- the number of nodes in our
multicomputer. In previous work, we have been concerned with
the topology and richness of the interconnection network
between the nodes of a multicomputer using CSS codes, finding
that a linear network is adequate for many
purposes~\cite{van-meter07:_commun_links_distr_quant_comput,van-meter07:_distr_arith_jetc}.
The extension of nodes into the serpentine teleportation
waveguide in Fig.~\ref{fig:qmc} enables such a linear-network
multicomputer, although the additional necessary resources for
bridging lossier chip-to-chip connections will not be
considered here.

\label{sec:fabrication}

The structures in our architecture are large by modern VLSI
standards; the principle fabrication difficulty is
accurate creation of the gap between the cavities and the
waveguides. That spacing must be 10-100nm, depending on the
microdisk and waveguide size and quality
factors~\cite{koseki09:mono-integ}. The roughness of the cavity edge
is a key fabrication characteristic that determines the quality of
the cavity, and ultimately the success of our device.

Although the device architecture and quantum dot technology are not
yet fixed, we include images of test-devices fabricated using e-beam
lithography following the methodology described in
Ref.~\refcite{koseki09:mono-integ}, only to help visualize future
devices. Figures~\ref{fig:local-cz} and \ref{fig:cnot-purification}
include scanning electron microscope images of a device created in a
GaAs wafer containing a layer of self-assembled InAs quantum
dots~\cite{koseki09:mono-integ}.  More scalable fabrication techniques
than e-beam lithography must ultimately be developed for scalability;
promising routes include nanoimprint
lithography~\cite{StephenY.Chou04051996} and deep sub-wavelength
photolithography~\cite{LinjieLi05152009,TrishaL.Andrew05152009,TimothyF.Scott05152009}.


\subsection{Qubit Roles and Basic Circuits}
\label{sec:qubit-roles}

\begin{figure}
\includegraphics[width=\columnwidth]{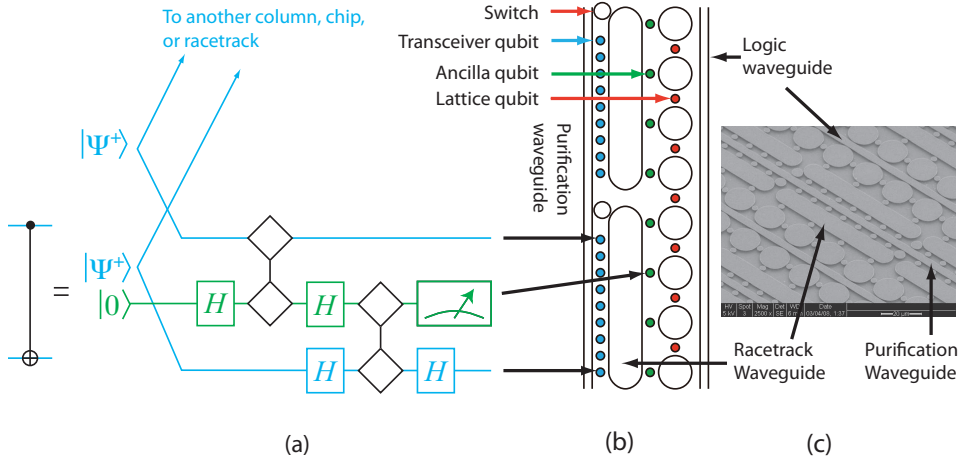}
\caption{(a) Partial circuit for executing purification on
  long-distance Bell pairs. The diamonds represent a probabilistic
  parity gate which projects two qubits into an odd-parity subspace
  with probability of approximately 50\%. These gates are achieved via
  pulses routed through the racetrack waveguides via the
  ring-waveguide labelled ``switch''.  All measurements are in the $X$
  basis. (b) The basic layout unit is a column of racetrack and
  circular waveguides sandwiched between the straight purification and
  logic waveguides. (c) Zoom-out of the same device shown in
  Fig.~\ref{fig:local-cz}(b).}
\label{fig:cnot-purification}
\end{figure}

The different colors for the qubit quantum dots in
Figure~\ref{fig:cnot-purification} represent different roles within
the system.  Physically, the cavities are identical, but they are
coupled to different waveguides, allowing them to interact directly
with different sets of qubits.  Within those connectivity constraints,
their roles are software-defined and flexible.  Finding the correct
hardware balance among the separate roles is a key engineering
problem.  The answer will depend on many parameters of the physical
system, including the losses in switches and couplers, and will no
doubt change with each successive technological generation.

The red qubits in the figures, in the column vertically placed between
the larger circles, are the \emph{lattice} qubits.  Those that are
functional are assigned an effective $(x,y)$ position in the 2-D
lattice used to implement tQEC.  These are subsequently divided into
\emph{code} qubits, which are never directly measured, and
\emph{syndrome} qubits, which are regularly measured following
connections to code qubits in order to maintain the topologically
protected surface code. The ideal number and density of syndrome
qubits among code qubits depends on the yield.  Within a column, all
functional nearest neighbor pairs of qubits can be coupled in
parallel.  Non-nearest-neighbor couplings can only occur sequentially.
For very low yields, in which code qubits rarely have nearest-neighbor
couplings, only a few syndrome qubits per column are required as the
syndrome circuits must largely be implemented sequentially, implying
the syndrome qubits can be reused.

The blue qubits, or \emph{transceiver} qubits, are aligned with the
racetracks and the long purification waveguides.  These qubits are
used to create Bell pairs between column groups within the same
device, or between devices.  Because purification is a very
resource-intensive process, the transceiver qubits are numerically
the dominant type.

The green qubits, sandwiched between the column of circles and the
column of racetracks, are \emph{ancilla} qubits, used to
deterministically connect stochastically created entangled states
among (blue) transceiver qubits to (red) lattice qubits. The green
qubits also play an auxiliary role during the purification of the
blue qubits.

\begin{figure}
\centering
\includegraphics[width=\textwidth]{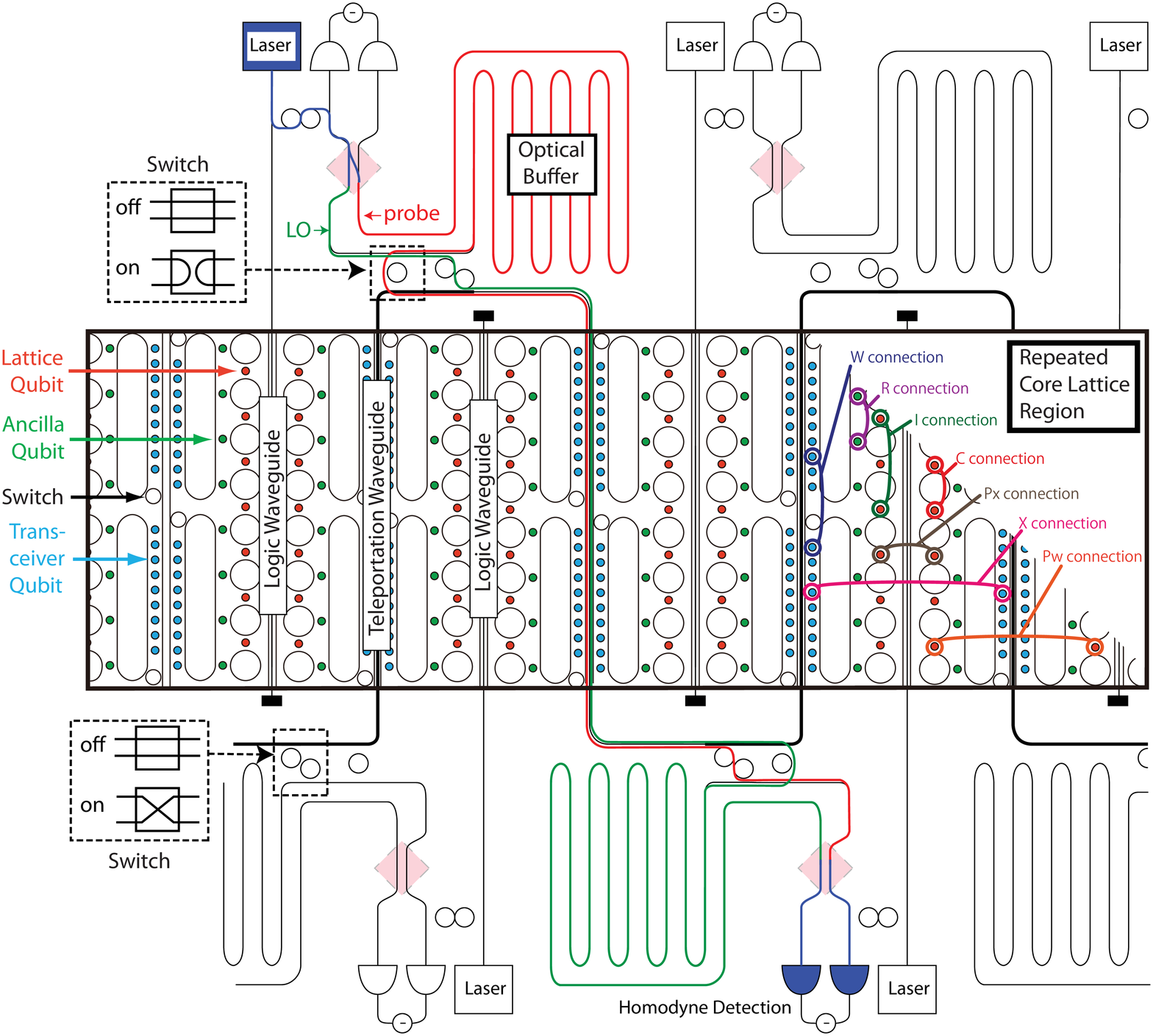}
\caption{The nanophotonic quantum multicomputer architecture.  Small
  microdisks containing lattice, ancilla, and transceiver qubits are
  color-coded while waveguides and microdisk-based add-drop switches
  are indicated by black lines.  This schematic indicates the critical
  elements of the nanophotonic chip-layout described in the text, but
  the structures shown are not to-scale.  In particular, the modulated
  CW lasers and detectors shown are the largest elements and are
  likely to be off-chip.  The pink squares indicate the location of
  beam-splitters defined by evanescently coupled ridge-waveguides,
  which split a single laser pulse (indicated by a blue line) into
  probe (red line) and local oscillator (LO, green line) optical pulses.
  These pulses travel two paths; one is buffered by a serpentine
  waveguide which delays the probe by several times the pulse width of
  approximately 100~ps.  (The pulse colors are schematic only; these
  pulses are to be monochromatic.)  The probe is switched to follow
  the LO along the same route through the teleportation waveguides of
  the core chip, which depend on the qubits to be coupled.  Single
  passes from top-to-bottom, such as the one shown by the red and
  green lines, enable the similar ``W connections'' and ``Pw
  connections'' between qubits as shown on the right.  A U-shaped path
  (not-shown) would enable the longer-distance ``X'' and ``Px''
  connections.  Lasers directly coupled into waveguides enable C
  connections and mediate logic within the circular microdisks
  connecting lattice qubits to ancilla qubits.  The rectangular region
  in the center is repeated many times vertically and horizontally.}
\label{fig:qmc}
\end{figure}

The circuit, or program, for executing purification on the blue
qubits is shown in Figure~\ref{fig:cnot-purification}.   The
blue qubits have previously been measured and are thus
initialized to a known state. Then, qubits in a given
teleportation column of Figure~\ref{fig:qmc} are entangled with
qubits in either the same column or the one neighbouring it to
the right using the heralded entanglement generation technique
discussed in Sec.~\ref{sec:gates}. Note that waveguide loss
prevents the efficient entangling of qubits in widely separated
teleportation columns.  In general, a laser pulse is inserted
in the teleportation waveguide at a given column, coupled with
a qubit in that column, coupled with a second qubit either in
that column or the one neighbouring it to its right and then
switched out of the teleportation waveguide and measured.  This
process is repeated in rapid succession, building a pool of
low-fidelity entangled pairs, creating the $|\Psi^{+}\rangle$
states at the left edge of Figure~\ref{fig:cnot-purification}.

\begin{figure}
\includegraphics[width=\columnwidth]{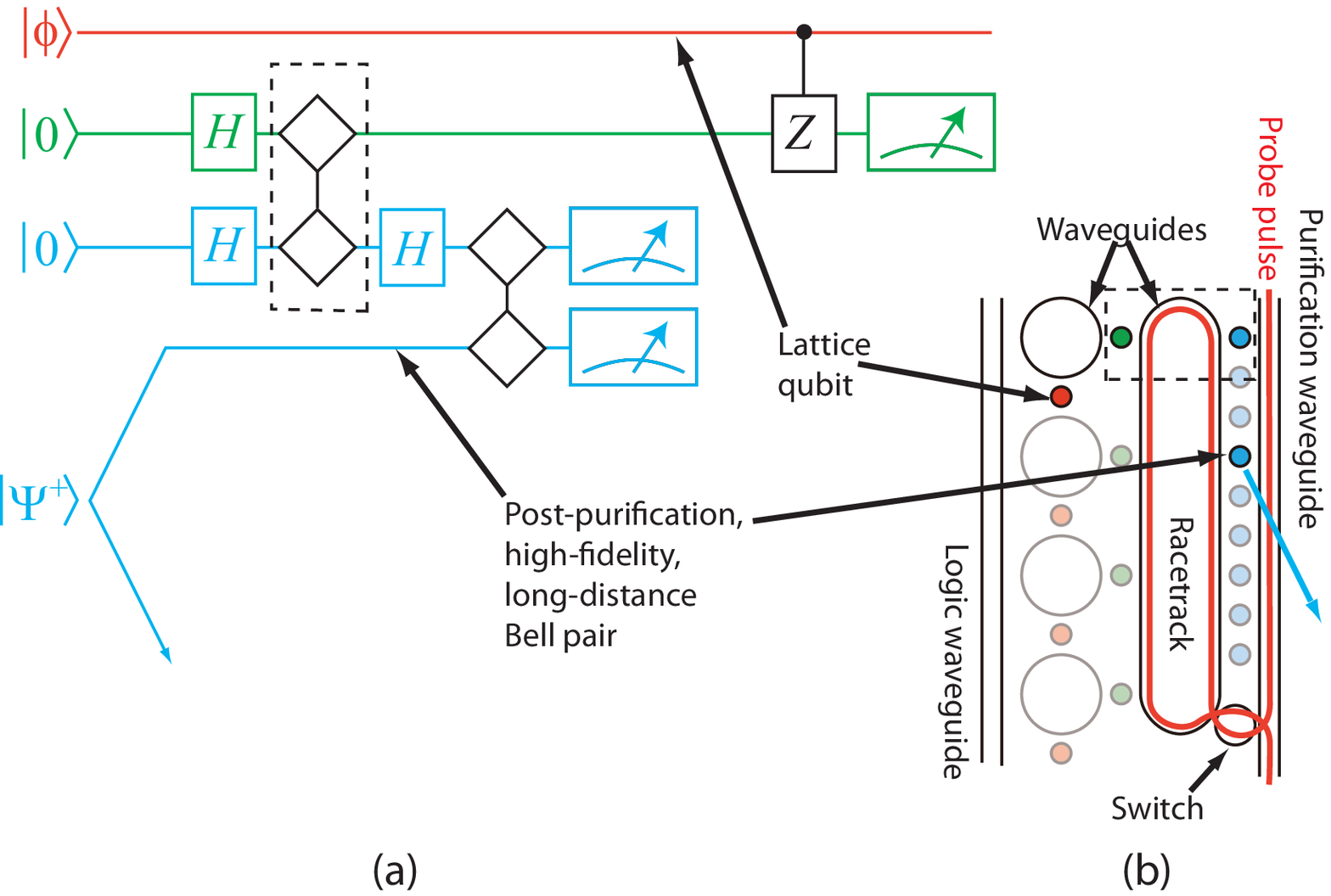}
\caption{(a) Partial circuit and (b) qubit/cavity layout and pulse
  path for executing long-distance clustering operations.  This
  circuit and a matching one elsewhere in the system execute the
  logical controlled-Z gate between two lattice (red) qubits in a
  teleported fashion (which we call telegate) by using a high-fidelity
  Bell pair built on transceiver (blue) qubits. The four qubits used
  in this circuit are highlighted in the layout. The second
  transceiver qubit and the ancilla (green) are used as ancillae in
  this circuit. The diamonds represent probabilistic ($P \approx 50\%$)
  parity gates on the racetrack-shaped waveguide, between either the
  two transceiver qubits or the transceiver and the ancilla.  The gate
  in the dashed-line box in (a) is executed by enabling the two qubits
  in the box in (b).  All measurements are in the $X$ basis.  The
  physical CZ gate in the top row is performed using the circuit of
  Figure~\ref{fig:local-cz}.}
\label{fig:cnot-cluster}
\end{figure}


Once the base-level entangled pairs are created, the circuit in
Figure~\ref{fig:cnot-purification} is executed within each
column, which employs two probabilistic parity gates to achieve
the controlled-NOT operations used in entanglement
purification. Purification proceeds until entangled state
fidelities are considered sufficient for computation.  At that
time the purified entanglement between blue transceiver qubits
is used to make an appropriate entangled (green) ancilla which
are connected to the target lattice qubits.

Finally, the high-fidelity Bell pairs are used to create the tQEC
lattice, using the clustering circuit shown in
Fig.~\ref{fig:cnot-cluster}.

\subsection{Lattice}
\label{sec:lattice}

The most important issue in the generation of a cluster state in our
geometry is the physical asymmetry between connections within a
column, those with other columns, and those between dies.  The
hierarchy of connection distances in our system will be
characterized in terms of the number of laser pulses and
measurements required to achieve entanglement of a particular
fidelity.

Entangling two qubits connected to the same circular waveguide is
straightforward; we can refer to these as ``cavity connected'' or
``C-connected.''  Racetracks are a longer, and slightly
lower-fidelity, form of cavity; we refer to two ancillae or two
transceiver qubits on the same racetrack as ``R-connected'', or
racetrack-connected.  Two lattice qubits connected through an
R-connected Bell pair are said to be indirectly connected, or
``I-connected''.

Within a logic column, many deterministic gates on C-connected
qubits can be performed without purification, and a high level of
parallelism may be employed. The pulses that execute deterministic
gates on the logic waveguide couple into the cavities only weakly,
and do not need to be measured after the gate, making it possible
that the same strong pulse could be used to execute several gates
concurrently.  If we label the qubits with the pattern $ABABA...$,
we may be able to couple all of the $AB$ pairs in one entangling
time slot, then couple all of the $BA$ pairs in the second time
slot.

The fidelity of W connections is dominated by the efficiency of
coupling pulses into and out of cavities, as the loss in the
waveguide will be negligible.  When connecting two lattice qubits in
columns separated by a purification waveguide, we require moderate
amounts of purification.  The purification ancillae are themselves
W-connected; the post-purification lattice connection we refer to as
``$P_W$-connected''.

Finally, qubits that do not share the same purification waveguide
must be connected using a pulse that transits one or more switches.
We refer to these physical connections as $X$ or $X_{i,j}$
connections, where $i$ is the number of switches and $j$ is the
number of I/O ports that must be transited. Lattice qubits connected
after purification we refer to as $P_X$-connected.

The $P_W$-connections and $P_X$-connections will be most strongly
subject to bottlenecks from the limited number of laser pulses and
detection events in our architecture, and are therefore the focus of
our numeric studies in the next section.

\section{Resource Estimates}
\label{sec:results}

Given a set of technological constraints (pulse rate, error rate,
qubit size, maximum die size), a complete architecture will balance a
set of tradeoffs to find a sweet spot that efficiently meets the
system requirements (application performance, success probability,
cost).  Minimizing lattice refresh time is the key to both
application-level performance and fault tolerance, but demands
increased parallelism (hence cost); in our system, this favors a very
wide, shallow lattice, which is more difficult to use effectively at
the application level.  Increasing the number of application qubits
increases the parallelism of many applications (including the modular
exponentiation that is the bottleneck for Shor's algorithm), but if
the space dedicated to the singular factory does not increase
proportionally, performance will not improve.

We begin by describing the communication costs and the impact of loss
on the lattice refresh cycle time in a generic 2-D multicomputer
layout, from which we can calculate the effective logical clock cycle
time for executing gates on application qubits.  With these concepts
in hand, we then propose an architecture, and calculate its
prospective performance.

\subsection{Communications and Lattice Refresh}
\label{sec:comm-cost}

\begin{figure}
\centering
\includegraphics[width=8cm]{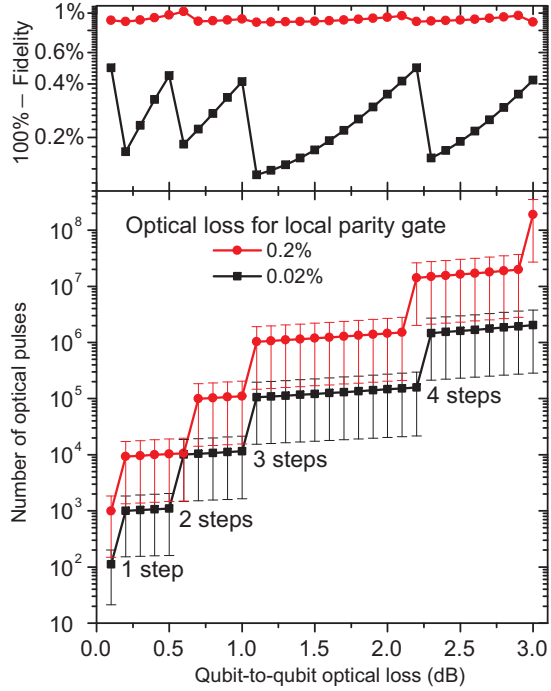}
\caption{For qubus connections, impact of signal loss on the final
  fidelity achievable using symmetric purification.  Error bars
  represent the RMS of the number of pulses, which is close to the
  average number; the distribution is strongly Poisson-like.}
\label{fig:finfid}
\end{figure}

Figure~\ref{fig:finfid} shows the residual infidelity and the
cost in teleportation waveguide pulses as a function of the
loss in the probe beam from qubit to qubit through the
waveguides.  Purification is performed using only Bell pairs of
symmetric fidelities, and is run until final fidelity saturates
or until fidelity is better than 99.5\%.  The two curves
represent two values of round-trip loss in the racetrack
waveguides used for local parity gates; with local loss of
0.2\%, we cannot achieve a final fidelity above the threshold
for tQEC. Thus, we establish an engineering goal of 0.02\% loss
or better.

The values in Fig.~\ref{fig:finfid} are calculated by
generating a Markov probability matrix for the protocol of
symmetric purification\cite{dur:PhysRevA.59.169}, where each
matrix transition requires the generation and detection of an
optical pulse in the teleportation waveguide. Probabilities and
fidelities for each step are found using the formalism
presented in Ref.~\refcite{hybridnjp}. Many of these transitions
are deterministic, but some are not due to the probability of
parity gates failing or the purification protocol failing.
Exponentiation of this matrix allows the direct calculation of
the probability of completing the protocol in a given number of
steps, allowing calculation of the probability density function
for completion of purification vs. number of optical pulses.
These probability distributions are strongly Poissonian.  They
are used to calculate the average and root-mean-square number
of pulses plotted in Fig.~\ref{fig:finfid}.

This Markov analysis is useful for estimating performance, but
overestimates the required spatial and temporal resources
considerably.  The strictly symmetric purification routine assumed
here makes less than ideal use of qubit memory; alternative resource
management strategies can lead to order-of-magnitude improvements in
speed without a comparable increase in size, as considered, for
example, in Ref.~\refcite{van-meter07:banded-repeater-ton}.  Also, the
calculation we have performed assumes that when parity gates fail in
the circuit shown in Fig.~\ref{fig:cnot-purification}(a), the entire
procedure fails and entangled pairs must be regenerated and
repurified.  In fact, if one parity gate succeeds and the other fails,
then one Bell pair preserves some of its entanglement and may be kept,
possibly with a Pauli correction, for subsequent purification rounds.
Optimizing the purification procedure to account for such
possibilities is difficult to do analytically; Monte Carlo simulations
such as those in Ref.~\refcite{van-meter07:banded-repeater-ton} may
estimate the worth of these strategies, but we leave such simulations
for future work.

With the proper layout, we can connect multiple chips into a
two-dimensional structure.  With $V$ rows of $H$ chips each, and a
chip that consists of $C$ columns each containing $R$ rows of lattice
qubits, we have a physical structure capable of supporting an $HC
\times VR$ lattice.  In such a multicomputer, entangling pulses may be
destined for another qubit in the same column in the same chip,
another qubit in the same column but the chip below, or in the
neighboring column to the left or right.  With multiple possible
destinations, switching is naturally required; we can arrange the
switching so that vertical connections are $X_{1,1}$ connections and
horizontal ones are $X_{2,1}$ connections.  Assessing the scalability
of such a system and establishing guidelines for configuring the
system depend on understanding these connections.

Table~\ref{tab:big-lat-costs} lists the costs for the lattice building
operations on such a switched multicomputer architecture.  We compare
two logical lattices, a direct-mapped $HC\times VR$ logical lattice
and a sub-lattice-organized $HCs \times VR/s$ logical lattice in which
each physical column is used as a small $R/s\times s$
lattice~\footnote{The table assumes that $R \bmod s = 0$.  Although
  that is not a requirement, the expressions are more complex for $R
  \bmod s \ne 0$; without careful structuring, potentially as many as
  half of the $P_W$ connections may become $P_X$ for $X_{1,1}$.}.  The
physical yield affects the probability that two neighboring lattice
qubits and their shared ancilla are good, and hence the probability
that a $C$ connection can be used.  Additionally, for low yields ($y
<0.8$), we assign only a few qubits per column as tQEC syndrome
qubits, forcing all lattice cycle operations to use $P_W$-connected
gates.

\begin{table*}
  \tbl{Number and types of connections per physical waveguide for
    lattice-building
    for an $H\times V$ multicomputer with $C\times R$
    lattice qubits per node and $HC$ total laser input ports and
    lattice sub-factor $s$.
    Expressions assume $R \bmod s = 0$.  $R_{f} = Ry_e = Ry_p(1-(1-y_p)^2)$, the
    functional number of qubits in a column.}
  {
\begin{centering}
\label{tab:big-lat-costs}
\begin{tabular}{|p{1.2in}||p{0.75in}|p{2.1in}|}\hline
Connection type &  100\% yield & physical yield $y_p$ \\
\hline
$C$ & $2V(R-s)$ & $n_C = 2V(R_{f}-s)y_p^2$ (for $y_p \ge 0.8$) or 0
($y_p < 0.8$) \\
$P_W$ & $V(2R-R/s)$ & $n_W = V(2R_{f}-R_{f}/s)+2V(R_{f}-s) - n_C$ \\
$V$ neighbor ($P_X(X_{1,1})$) & $2s(V-1)$ & $n_{X1} = 2s(V-1)$ \\
$H$ neighbor ($P_X(X_{2,1})$) & $VR/s$ & $n_{X2} = VR_{f}/s$ \\
\hline
\end{tabular}
\end{centering}
}
\end{table*}

We observe several qualitative facts about this architecture:

\begin{itemize}
\item The lattice cycle time is constant as $H$ increases, but the
  number of lasers and measurement devices must increase
  proportionally.
\item To first order, the lattice cycle time scales linearly with $VR$,
  but second-order effects will likely make it worse than
  linear.
\item The number of $X_{2,1}$ connections favors a sub-lattice with a
  large $s$, but the minimum size of the logical lattice limits
  $s$; we require $14d \le VR/s$.
\item Increasing lattice cycle time hurts fidelity due to memory
  degradation.
\item Increasing lattice cycle time hurts application performance.
\end{itemize}

The total lattice refresh cycle time is $t_{lat} = t_{pulse}p_{lat}$,
where $p_{lat}$ is the number of pulse time steps in the complete
cycle.  The final, logical clock rate for application gates depends on
both the refresh cycle and the temporal extent of the lattice holes as
they move through the system to execute logical gates.  We can
visualize the movement of the holes through the temporal dimension as
``pipes'' routed in a pseudo-3-D space.  To maintain the same $4d$
perimeter and spacing about the hole as it extends into the temporal
dimension, each hole movement will also have to extend for $5d$
lattice refresh cycles.  We have used $d = 14$ as the length of one
side of each square hole.  The temporal spacing must be $4d = 56$,
implying that the fastest rate at which hole braiding can occur is $5d
= 70$ lattice refresh cycles.

In our architecture, the logical clock rate is $\Omega(d^2)$.
The number of refresh cycles per logical gate is $\Theta(d)$.
The refresh time itself is $\Omega(R) = \Omega(d)$; because we
must choose $R \propto d$, the number of pulses grows at least
linearly in $d$.  As the columns lengthen, fidelity falls and
the number of pulses per cycle grows, creating a positive
feedback in $d$ and cycle time.

\subsection{Proposed Architecture and Performance}

Table~\ref{tab:cand-1} summarizes our initial strawman
architecture, depicted in Fig.~\ref{fig:qmc}.  To factor an
$n$-bit number using Shor's algorithm, we would like to have
$6n$ logical qubits. Having established a goal of factoring a
2,048-bit number, we need 12,288 logical qubits.

\begin{table*}
  \tbl{Summary of our proposed serpentine, add-drop filter
    architecture. M$ = 2^{20} \sim 10^6$.}
{
\begin{centering}
\label{tab:cand-1}
\begin{tabular}{|p{2in}|p{2.5in}|}\hline
\emph{System Hardware} & \\
Chip lattice, $C\times R$ & $128\times 770$ \\
Multicomputer setup, $H\times V$ & $65536\times 1$ \\
Physical lattice size (in qubits) & 8M$\times 770 = 6.46\times 10^{9}$ \\
Laser ports & $4$M \\
Measurement devices & $16$M \\
Purification/entanglement pulse rate & 10~GHz \\
Switch type & add-drop filter \\
Required physical yield & $y_p = 40\%$ \\
Effective yield for lattice qubits & $y_e =
y_p(1-(1-y_p)^2) = 25.6\%$ \\
Functional column height & $R_{f} = Ry_e = 196$ \\
Required local optical loss & 0.02\% \\
Required adjusted gate error rate & $p_{err} \le p_{thresh}/4 \sim 0.2$\% \\
Required memory coherence time & $t_{mem} \ge 1000t_{lat} = 49$~msec \\
\hline
\emph{Communication Costs} & \\
$W, P_W$ connection & 0.1dB, $p_W = 111$ pulses \\
$X_{0,0}, P_X$ conn. (neighboring column) & 0.4dB, $p_{X} = 1068$ pulses \\
\hline
\emph{Lattice Operations} & \\
Sub-lattice factor $s$ & 1 \\
Logical lattice & 8M $\times 196$ \\
Pulses per lattice cycle (avg.) & $p_{lat} \sim n_Wp_W + n_{X2}p_X = 4.9\times 10^{5}$ \\
Lattice cycle time &  $t_{lat} = p_{lat}t_{pulse} = 49~\mu$sec \\
\hline
\emph{Logical Qubit Operations} & \\
Hole separation constant & $d = 14$ \\
Lattice area per qubit (at rest, loosely packed) & $14d\times 9d =
196\times 126 = 24696$\\
Lattice area per qubit (at rest, tightly packed) & $10d\times 5d =
140\times 70 = 9800$\\
Hole movement time & $t_{move} = 5dt_{lat} = 3.41$~msec \\
Hole braiding time & $t_{braid} = 5dt_{lat} = 3.41$~msec \\
Toffoli gate construction & Nielsen \& Chuang~\cite{nielsen-chuang:qci}, p. 182 \\
Finished $|S\rangle$ states per Toffoli gate (avg.) & 11.5 \\
Total braidings of $|S\rangle$ states per Toffoli & $1795$ \\
Toffoli gate time $t_{tof}$ & $\sim 14t_{braid} = 48$~msec \\
\hline
\emph{Application Operations} & \\
Maximum capacity, in logical qubits & 119836 \\
Number of application logical qubits & $6n = 12288$ \\
$|S\rangle$ factory space & 77589 \\
``wiring'' space & $25\% = 29959$ \\
\hline
\emph{Shor} & \\
Length of number to be factored & $n = 2048$ \\
Adder & Carry-lookahead \\
Adder time & $t_{add} = 4\log_2{n}t_{tof} = 2.1$ seconds \\
Modulo \& indirect arithmetic & $w = 2, p = 11$, $\sim 5\times$
faster
than basic VBE~\cite{vedral:quant-arith,van-meter04:fast-modexp} \\
Number of adder calls & $n_{add} = 4n^2 = 1.68\times 10^{7}$ \\
Number of adders executed in parallel & 1 \\
Number of Toffoli gates & $n_{tof} = 40n^3 = 3.2\times 10^{11}$ \\
Time to execute algorithm only & $3.5\times 10^{7}$ seconds (409 days)
\\
Time to create singular states & $2.7\times 10^{7}$ seconds (314 days)
\\
Final execution time & 409 days \\
\hline
\end{tabular}
\end{centering}
}
\end{table*}

\begin{figure}
\centering
\includegraphics[width=12cm]{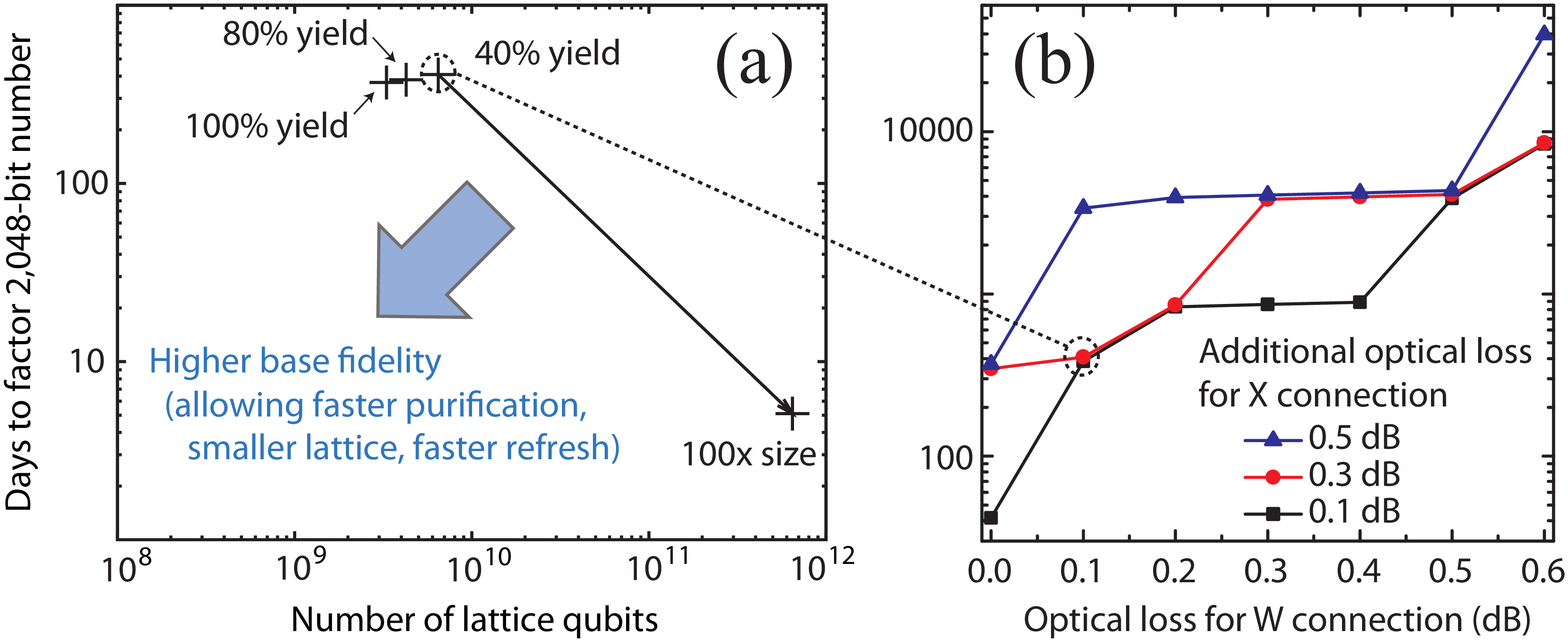}
\caption{Factoring time for 2,048-bit number using Shor's factoring
  algorithm. a) Our baseline proposal, with 40\% yield, 0.1dB W
  connections and 0.4dB X connections, can be improved by increasing
  the size and application-level parallelism of the system. Improving
  yield above 40\% reduces necessary resources only moderately, but
  raising the fidelity of the base-level entangled pairs has a major
  impact on both system size and performance. b) Achieving low-loss
  connections is critical to performance.}
\label{fig:factoring-time}
\end{figure}

Ultimately, the execution of application algorithms in tQEC requires,
as at the physical level, two components: communication and
computation. Logical communication consists of routing the pipes
through the pseudo-3-D lattice.  These pipes can route through the
space with only a fixed temporal extent, allowing the equivalent of
``long distance'' gates in the circuit model.  They do, however,
consume space in the lattice, creating a direct tradeoff between the
physical size of the system and the time consumed.  Additionally, the
shape of the logical lattice determines how efficiently logical qubits
can be placed and routed.  We assign 25\% of the logical qubit space
for wiring and hole movement space.

Computation, for many algorithms, will be dominated by Toffoli
gates; as some of the operations are probabilistic, an average
of over ten $S$ and $T$ states are required for each. Shor's
algorithm requires some $40n^3$ Toffoli gates: $5n^2$ adder
calls~\cite{vedral:quant-arith} (after optimizations to modulo
arithmetic and one level of indirection in the
arirthmetic~\cite{van-meter04:fast-modexp}), each requiring
$10n$ Toffoli gates~\cite{draper04:quant-carry-lookahead}.  The
total of $40n^3 = 3.2\times 10^{11}$ Toffoli gates require over
$10^{12}$ $S$ states.  Again, a direct tradeoff can be made
between space and time, as the $S$ states can be built in
parallel.  For our system and this size of problem, rough
balance is achieved with about 65\% of the logical qubits
dedicated to the $|S\rangle$ factory.

The multicomputer organization is wide and shallow, to minimize
refresh cycle time.  Once we have decided to limit $V$ to 1, the
detailed chip layout simplifies, allowing the serpentine waveguide
shown in Fig.~\ref{fig:qmc}.  In this architecture, W connections are
high fidelity, there are no $V$ neighbors ($X_{1,1}$ connections), and
connections to neighboring columns need not leave the chip except at
chip boundaries.  The $n_{X2}$ from Table~\ref{tab:big-lat-costs} is
still $VR_f/s$, but physical connections are $X$ connections with a
loss of only about 0.4dB.  The vertical height of a single chip will
only accommodate enough cavities for a direct-mapped lattice, $s =
1$.

Figure~\ref{fig:factoring-time}a shows the execution time for our
proposed system.  A 2048-bit number should be factorable in just over
400 days, if the technological characteristics in
Table~\ref{tab:cand-1} can be met.  The system is large, requiring
more than six billion lattice qubits and several times that total
number when ancillae and transceivers are included.  At the
application level, much more parallelism is available if a larger
system is built.  A system one hundred times larger would factor the
number in about five days.

Figure~\ref{fig:factoring-time}b shows execution time as a function of
the loss in our two key connection types, the intra-column W
connections and the inter-column X connections.  Minimizing the
additional loss incurred in inter-column travel helps hold execution
time within reasonable bounds.

Reaching toward the desirable lower left corner of
Fig.~\ref{fig:factoring-time}a requires improving the base-level
entanglement fidelity or reducing the number of pulses used to purify
Bell pairs.  Our system is fairly robust to yield.  Below 40\% it is
difficult to build a system capable of running tQEC, but above that
level, increasing yield has only minor effects on temporal and spatial
resources.  This gives a clear message: pursue fidelity and quality of
components at the expense of yield.

\section{Discussion}
\label{sec:conclusion}

Our design focuses on the communications within a quantum
computer, building on a natural hierarchy of connectivity
ranging from direct coupling of neighbors on one physical axis
of our chip through medium-fidelity, waveguide-based
purification coupling on the other axis, to distant, switched
connections requiring substantial purification.  Thus, while we
refer to our design as a quantum multicomputer with each node
consisting of a single chip, it is more accurate to regard the
connections between qubits as occurring on a set of levels
rather than a simple internal/external distinction. Founded on
quantum dots connected via cavity QED and nanophotonic
waveguides and using topological error correction, this
proposal represents progress toward a practical quantum
computer architecture. The physical technologies are maturing
rapidly, and tQEC offers both operational flexibility and a
high threshold on realistic architectures such as ours.

While the overall architecture (multicomputer) and the system
building blocks (tQEC, purification circuits, etc.) have been
established, much work remains to be done.  The most important
pending decision is the actual choice of semiconductor and
quantum dot type.  The cavity $Q$ and memory lifetime, which
dramatically affect our ability to build and maintain the
lattice cluster state, will be critical factors in this
decision. The yield of functional qubits will ultimately drive
the types of experiments that are feasible.

With the decision of semiconductor and the key technical
parameters in hand, it will become possible to more
quantitatively analyze the mid-level design choices of node
size, layout tradeoffs, and the numbers of required lasers and
photodiodes.  The control system for managing the qubits and
cavity coupling will be a large engineering effort involving
optics, electronic circuits, and possibly micromechanical
elements.  Finally, application algorithms need to be
implemented and optimized and run-time systems deployed, which
will require the creation of large software tool suites.

One of our goals in this work is to establish target values for
experimental parameters that must be achieved for such a large system to
work.  For the chip design and system configuration we present here, we
estimate that the yield of functional quantum dots must be at least
$40\%$, the local optical loss must be better than $0.02\%$, the
adjusted gate error rate better than $0.2\%$, and the memory coherence
time about $50$ milliseconds or more.  The exact values of these goals
depend on the architecture, system scale, and application; the entire
system is summarized in Table~\ref{tab:cand-1}.

As a final comment, the physical resources demanded by this
architecture are daunting.  Other architectures for quantum computers
are comparably daunting.  The current work is intended in large part
to reveal the scope of the problem.  With realistic resources such as
lossy waveguides, finite-yield qubits, and finite chip-sizes, the
added overhead for error correction makes quantum computers very
expensive by current standards. We must rely on engineering
advancements to improve nanophotonic and quantum dot devices as well
as VLSI-like manufacturing capabilities to realize a quantum computer
with a realistic cost.  Indeed, our current understanding of how to
make very large quantum computers is often likened to classical
computers before VLSI techniques were developed.  The successful
technologies enabling practical approaches to building large computers
are likely yet to be discovered, but architectures such as the one we
have presented and the defect-tolerant, communication-oriented design
principles we have used are expected to provide the guiding context
for these new technologies.

\section*{Acknowledgments}

This work was supported by NSF, with partial support by MEXT
and NICT. We acknowledge the support of the Australian Research
Council, the Australian Government, and the US National
Security Agency (NSA) and the Army Research Office (ARO) under
contract number W911NF-08-1-0527. The authors thank Shinichi
Koseki for fabricating and photographing the test structure and
Shota Nagayama for help with the figures. We thank Jim
Harrington, Robert Raussendorf, Ray Beausoliel, Kae Nemoto,
Bill Munro, and the QIS groups at HP Labs and NII, for many
useful technical discussions.  We also would like to thank
Skype, Ltd. for providing the classical networking software
that enabled the tri-continental writing of this manuscript.

\bibliographystyle{unsrt}

\bibliography{paper-reviews,bibstrings,TLbib,quantumdots,diamond,silicon,qubus}

\end{document}